\documentclass{iaus}
\usepackage{graphicx,natbib}

\title[MHD and turbulence in SPH]{Magnetic fields and Turbulence in Star Formation using Smoothed Particle Hydrodynamics}

\author[Daniel Price] {Daniel J. Price}

\affiliation{Centre for Stellar and Planetary Astrophysics, \\ Monash University, Clayton Vic 3800, Australia.
\\ email: {\tt daniel.price@monash.edu}}

\pubyear{2010}
\volume{270}  
\pagerange{119--126}
\jname{Computational Star Formation}
\editors{Alves, Elmegreen, Girart, Trimble, eds.}
\begin{document}

\maketitle

\begin{abstract}
 Firstly, we give a historical overview of attempts to incorporate magnetic fields into the Smoothed Particle Hydrodynamics method by solving the equations of Magnetohydrodynamics (MHD), leading an honest assessment of the current state-of-the-art in terms of the limitations to performing realistic calculations of the star formation process. Secondly, we discuss the results of a recent comparison we have performed on simulations of driven, supersonic turbulence with SPH and Eulerian techniques. Finally we present some new results on the relationship between the density variance and the Mach number in supersonic turbulent flows, finding $\sigma^{2}_{\ln \rho} = \ln (1 + b^{2} \mathcal{M}^{2})$ with $b=0.33$ up to Mach~20, consistent with other numerical results at lower Mach number \citep{ls08} but inconsistent with observational constraints on $\sigma_{\rho}$ and $\mathcal{M}$ in Taurus and IC5146.

\end{abstract}

\firstsection 
\section{Introduction}
 Magnetic fields and turbulence are thought to be two of the most important ingredients in the star formation process, so it is critical that we are able to model the effect of both in star formation simulations. Smoothed Particle Hydrodynamics (SPH, for recent reviews see \citealt{monaghan05,price04}), since it is a Lagrangian method where resolution follows mass, is a very natural method to use in order to model star formation. However the introduction of magnetic fields in SPH has a somewhat troubled history, and there are perceptions that the explicit use of artificial viscosity terms in order to treat shocks is too crude to enable accurate modelling of supersonic turbulence. We will discuss both of these aspects in this talk.

\section{Magnetic fields in SPH}
\subsection{A prehistory of MHD in SPH}
 Magnetic fields were introduced in one of the founding SPH papers by \citet{gm77}, though a detailed investigation did not follow until around a decade later with the publication of Graham Phillips' PhD work \citep{pm85} (PM85), with successful application of the method to star formation problems involving the collapse of isothermal, non-rotating, magnetised clouds \citep[e.g.][]{phillips86a}\footnote{Graham Phillips now pursues a rather different career as presenter of the science show `Catalyst'  on Australian television (I suspect this was easier than getting MHD in SPH to work).}. Indeed, it was \citet{pm85} who first discovered that the equations of motion for MHD -- when expressed in a momentum-conserving form -- were unstable to a clumping or `tensile' instability whereby particles attract each other unstoppably along field lines, that occurs in the regime where the magnetic pressure exceeds the gas pressure (i.e., $\beta < 1$).
 
   Whilst Phillips \& Monaghan proposed a way of achieving stability, another decade followed before the matter was investigated in any more detail, by another PhD student of Joe Monaghan's, Joe Morris. Morris' PhD thesis \citep{morrisphd} contained detailed stability analysis of the equations of the now-named Smoothed Particle Magnetohydrodynamics (SPMHD) and proposed a way of avoiding the instability by using more accurate derivatives for the anisotropic part of the magnetic force in place of exact conservation of momentum. Around the same time \citet{meglicki95} applied a non-conservative SPMHD formulation to magnetic fields in the Galaxy. The period after these works saw a few but limited applications of SPMHD to `real' astrophysical problems, including notably an application to magnetic fields in galaxy clusters by \citet*{dbl99}\footnote{Harald Lesch is now also a science show presenter, on German television.} and at least one foray into star formation by \citet{bp96}. However it would be fair to say that there was no real consensus on a standard approach and that many problems remained.

\subsection{21st century Smoothed Particle Magnetohydrodynamics}
 \citet*{bot01} suggested an alternative to Morris' stabilisation based on explicitly subtracting the non-zero source term (-${\bf B}\nabla\cdot{\bf B}$) that is the cause of the stability problems in the momentum-conserving formulation. They also proposed a regularisation scheme for the SPH particle distribution which, though complicated, was found to give excellent results on MHD shock problems by the focussing of resolution where it is needed. \citet{pm04a} adopted (but later abandoned) a stability fix proposed by \citet{monaghan00} and showed how the dissipative terms could be formulated for treating MHD shocks and other discontinuities in a standard SPMHD approach (that is, without regularisation). Paper~II \citep{pm04b} showed how the terms necessary for strict conservation in the presence of a spatially variable smoothing length could be included by deriving the SPMHD equations of motion from a variational principle. At the same time \citet{bot04} undertook a detailed stability analysis showing that indeed their source-term approach was indeed stable in more than one dimension. 
 
 With these improvements meant that SPMHD could be successfully run on many of the test problems used to test grid-based MHD codes \citep[Paper~III,][]{pm05}. However, accuracy in realistic problems was found to depend on the accuracy with which the divergence-free condition on the magnetic field -- perhaps the key problem for any numerical MHD scheme -- since the $\nabla\cdot{\bf B} = 0$ constraint enters the ideal MHD equations only as an initial condition:
\begin{equation}
\frac{\partial {\bf B}}{\partial t} = \nabla \times ({\bf v}\times {\bf B}); \hspace{1cm} \frac{\partial}{\partial t} \left( \nabla\cdot{\bf B} \right) = 0.
\label{eq:ind}
\end{equation}
 
  As aptly summarised by Mike Norman (this proceedings), approaches to the divergence constraint can be roughly categorised into 
 ``ignore'', ``clean'' or ``prevent''. Many test problems for MHD can be run quite happily (in SPH or otherwise) with the ``ignore'' approach: simply evolve the induction equation (\ref{eq:ind}) and cross your fingers (I refer to this as the ``hope and pray'' method).
For realistic applications this is almost never a good idea and in SPMHD simulations of star formation manifests as `exploding stars' due to large errors in the magnetic force (see \citealt{pf09} for a figure). \citet{pm05} examined a range of cleaning methods for the divergence constraint, including most promisingly the hyperbolic/parabolic cleaning scheme proposed by \citet{dedneretal02}. However this was not found to be sufficient to prevent divergence errors generated in realistic flows, least of all during protostellar collapse.

\subsection{The Euler Potentials}
 Thus, \citet{pb07} and \citet{rp07} adopted a preventative approach, based on writing the magnetic field in terms of the `Euler Potentials' \citep{stern76}, variously known as `Clebsch' or `Sakurai' variables:
\begin{equation}
{\bf B} = \nabla \alpha \times \nabla \beta,
\label{eq:eps}
\end{equation}
with which the divergence constraint is satisfied by construction. Furthermore the ideal MHD induction equation takes the particularly simple form
\begin{equation}
\frac{d\alpha}{dt} = 0; \hspace{1cm} \frac{d\beta}{dt} = 0,
\label{eq:epsind}
\end{equation}
corresponding physically to the advection of magnetic field lines by Lagrangian particles. This is then combined with a standard SPMHD approach by computing ${\bf B}$ according to Eq.~\ref{eq:eps} and using the standard equation of motion with either the Morris or \citet{bot01} force stabilisation (thus is is usually trivial to revert to a ${\bf B}$-based formulation -- that is, solving Eq.~\ref{eq:ind} instead of Eqs.~\ref{eq:eps}--\ref{eq:epsind} -- for comparison purposes.

 Adoption of this approach meant that we were able to study realistic star formation problems in 3D, including the magnetised collapse of a rotating core to form single and binary stars  \citep{pb07} and the effect of magnetic fields on star cluster formation from turbulent molecular clouds \citep{pb08}, most recently including both radiation and magnetic fields \citep{pb09}. Around this time magnetic fields were also implemented in the widely used SPH code GADGET \citep{ds09}.
 
 From these we have learnt that magnetic fields can significantly change star formation even with weak (supercritical) fields, significantly affecting the formation of circumstellar discs and binary stars (\citealt{pb07}, similar to the findings of many other authors, e.g.~\citealt{ht08,hc09,ml08,machidaetal08}) and having a profound effect on the star formation rate \citep{pb08,pb09}.
 
 However there are important limitations to the Euler Potentials (EPs) formulation. The most important are that:
\begin{enumerate}
\item They can only be used to represent topologically trivial fields,
\item As a consequence of a), the wind-up of a magnetic field in a rotating or turbulent flow cannot be followed indefinitely,
\item It is difficult to incorporate single fluid non-ideal MHD effects such as resistivity.
\end{enumerate}
 The topological restrictions are a consequence of the fact that the helicity ${\bf A}\cdot{\bf B}$ is identically zero in the EP representation. The lack of winding processes is easily understood as a consequence of Eq.~\ref{eq:epsind}: Since the potentials are simply advected by the particles, the reconstruction of the magnetic field via Eq.~\ref{eq:eps} relies on a one-to-one mapping between the initial and final particle positions. Thus, for example, a complete rotation of an inner ring of particles will return the magnetic field to its original configuration, whereas physically the field should `remember' the winding [in the talk this was illustrated by a dance with J. Monaghan which is difficult to insert in the proceedings].
 
  In practice this means that in using the EPs one is limited to the study of problems where the magnetic field is initially simple (e.g. imposed as a constant field) and at some scale the field winding will be lost (underestimated) in the calculations. The difficulty in representing physical resistive processes is also an issue given the importance of non-ideal MHD effects for star formation.

\subsection{Why don't you just use the Vector Potential?}
 At one of the last major computational star formation meetings, held at the Kavli Institute for Theoretical Physics during 2007, Axel Brandenburg suggested that the use of the vector potential, with the gauge set to give Galilean invariance, may provide similar advantages in terms of the divergence constraint without the topological restrictions of the Euler potentials. Three years and 25 pages of pain later \citep{price10} we had derived the Ultimate Vector Potential Formulation$^{\sc \rm TM}$ for SPMHD. It was beautiful, derived elegantly from a variational principle, the method was exactly conservative, with a brand new force equation that in principle might not suffer from instabilities since the divergence constraint was built-in... could it be that we had uncovered the ultimate method for MHD in SPH? Unfortunately the vector potential approach failed spectacularly in practice due to two problems: 1) The same old force-related instability in 2D and 3D, which could, however, be cured by reverting to the stable force formulations discussed above, though with more difficulty than in the standard SPMHD case, and 2) An instability peculiar to the 3D vector potential related to the unconstrained growth of non-physical components of ${\bf A}$, occurring, for example, when a 2D test problem like the Orszag-Tang vortex is run in 3D. Further investigation suggests the latter is related to not enforcing the gauge condition $\nabla\cdot{\bf A} = 0$ on the vector potential -- that is, simply pushing the divergence problem one level up. Thus, we must unfortunately report that the answer to the vector potential question is that it not a good approach for SPMHD.
 
 \subsection{Alternatives to the Euler Potentials}
 The most promising approach to lifting the restrictions of the Euler Potentials whilst preserving the benefits involve moving to a generalised formulation, where the most general involves three sets of potentials, in the form
\begin{equation}
{\bf B} =  \nabla \alpha_{1} \times \nabla \beta_{1} +  \nabla \alpha_{2} \times \nabla \beta_{2} +  \nabla \alpha_{3} \times \nabla \beta_{3},
\end{equation}
essentially taking advantage of the fact that, whilst sets of potentials cannot be added together, the fields can. Thus one can represent arbitrary field geometries. The reason for the three sets is because one of the potentials $\beta$ is essentially preserving one of the initial spatial coordinates so that a Jacobian map can be constructed -- for example 2D fields are represented by $\alpha = \alpha(x,y)$ and $\beta = z_{0}$. Thus in the most general case all three coordinates of the particles $(\beta_{1}, \beta_{2}, \beta_{3}) = (x_{0}, y_{0}, z_{0})$ are used to compute the mapping. Better still, adopting this more general formulation means that the potentials can be re-mapped to a new set whenever the gradients break down, meaning that physical field winding can be followed indefinitely, and with explicit control on the amount of numerical dissipation. Watch this space.

 An alternative is to return to the `clean' approach, despite the only limited success of the formulations considered by \citet{pm05}, at least for star formation problems. Use of cleaning methods has found more success in cosmological applications \citep{ds09} and there are further possibilities for improving the accuracy of the projection methods discussed in \citet{pm05} that offer some promise.
 
\section{Supersonic turbulence with SPH}
 Progress on solving the issues with magnetic fields in SPH generally slows due to the need to periodically address concerns over the core SPH method. Most recently this has involved a rather inflamed and unhelpful discussion of Kelvin-Helmholtz instabilities in SPH\footnote{Following the publication of \citet{agertzetal}. The issue itself has nothing to do with Kelvin-Helmholtz instabilities but rather to the (non-)treatment of contact discontinuities in standard SPH schemes, in this case manifested when a KH instability problem is run across a contact discontinuity. Adding a small amount of artificial conductivity is a simple fix (see \citealt{price08}).}, and which many authors are now using as the primary justification for the often fruitless pursuit of all manner of alternative particle methods.
\begin{figure}
\begin{center}
\includegraphics[width=0.45\columnwidth]{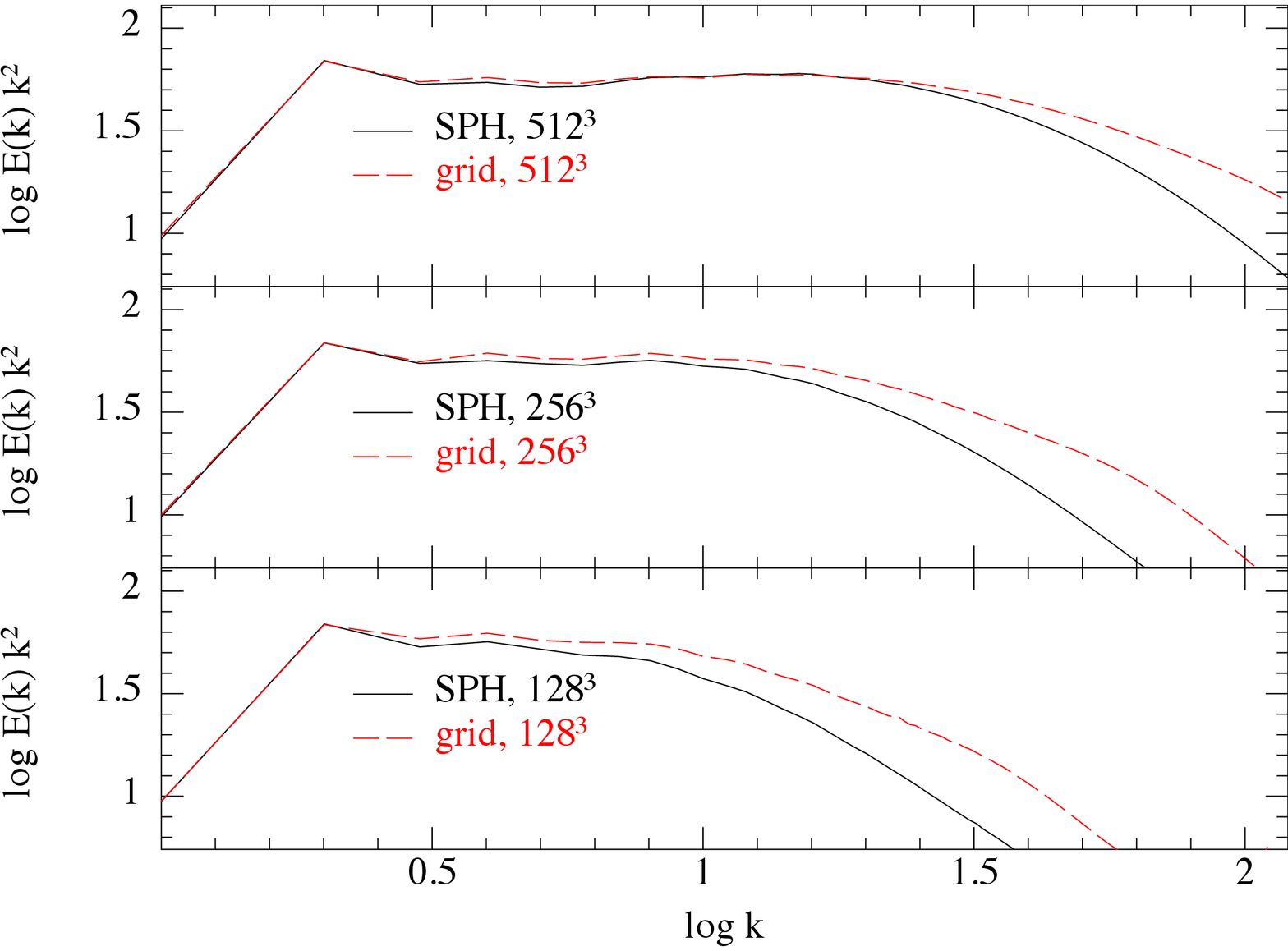} 
\includegraphics[width=0.45\columnwidth]{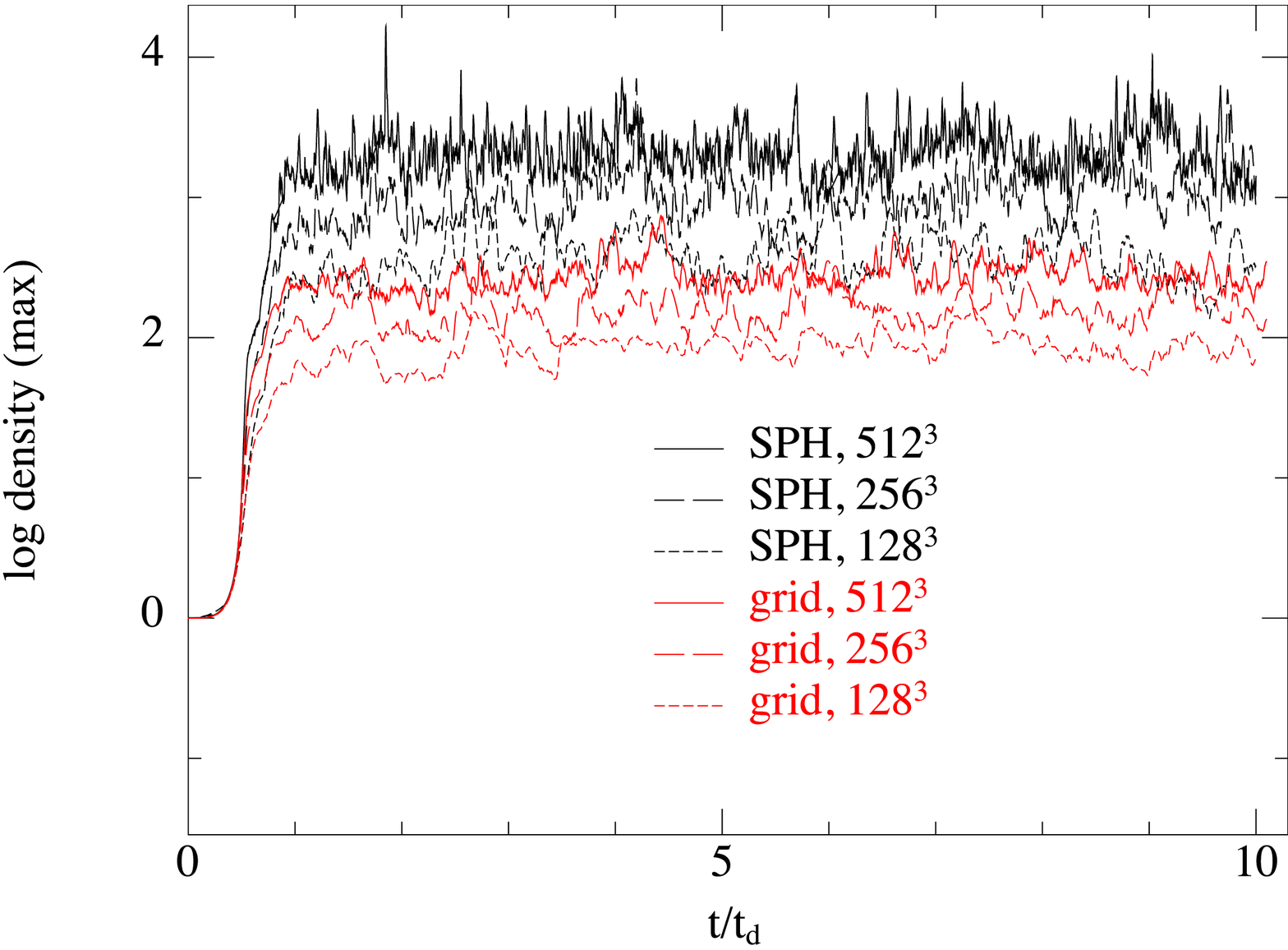} 
 \caption{Time-averaged kinetic-energy power spectra (left), and the time evolution of the maximum density (right) at three different resolutions in the SPH vs. grid driven turbulence comparison by \citet{pf10}. Comparable results on volume-weighted quantities such as the kinetic energy power spectrum requires roughly $n_{cells}\approx n_{particles}$ (left), but SPH resolves maximum densities at $128^{3}$ similar to a grid at $512^{3}$ (right).}
   \label{fig:turb512}
\end{center}
\end{figure}
 Computational star formation is no stranger to such debates, though thankfully at least the heated discussions surrounding `artificial fragmentation' \citep{kfm04} went quietly away, leading to a more helpful consensus on the implications of the physical approximations (e.g. use of a barotropic equation of state vs. radiation hydrodynamics) rather than arguments over numerics. However there remain sharp disagreements over numerics in some areas, for example the following in a discussion of turbulence simulations by \citet{bpetal06} given in \citet{padoanetal07}:
\begin{quote}
``SPH simulations of large scale star formation fields to date fail in all three fronts: numerical diffusivity, numerical resolution and presence of magnetic fields. This should case serious doubts on the value of comparing predictions based on SPH simulations with observational data (see also \citealt{agertzetal})'',
\end{quote}
(note the completely irrelevant reference to \citealt{agertzetal}). Such concerns over turbulence helped motivate at least two major comparison projects -- the `Potsdam' comparison \citep{kitsionasetal09} and the 2007 KITP comparison project, both studying the decay of supersonic turbulence from pre-evolved initial conditions.

 Given that comparisons of decaying turbulence are necessarily limited to a few snapshots and that there are issues surrounding the setup of initial conditions between codes, we additionally undertook a side-by-side comparison of driven turbulence, starting from simple initial conditions and over many crossing times, using just two codes -- D. Price's SPH code \textsc{phantom} and the grid-based \textsc{flash} code \citep{pf10}. The devastatingly obvious conclusion from such efforts was that to achieve comparable results between codes required -- wait for it -- comparable resolutions. Actually the point was rather more subtle, because ``comparable resolution'' depended on what quantity one was interested in. Thus, for example to obtain comparable power spectra in a volume-weighted quantity such as the kinetic energy required roughly equal numbers of SPH particles to Eulerian grid cells (and was thus about an order of magnitude more costly for SPH in terms of cpu time). On the other hand, the maximum density achieved in the SPH calculations at 128$^{3}$ particles was roughly equivalent to that reached in the grid-based calculations at 512$^{3}$ (thus being about an order of magnitude more costly for \textsc{flash}). Quantities involving a mix of the density and velocity fields -- such as the spectra of the mass weighted kinetic energy $\rho^{1/3} v$ -- showed intermediate results. Other quantities such as structure function slopes were not well converged in either code at $512^{3}$ computational elements. We refer the reader to \citet{pf10} for more details.

\begin{figure}
\begin{center}
\includegraphics[width=0.8\columnwidth]{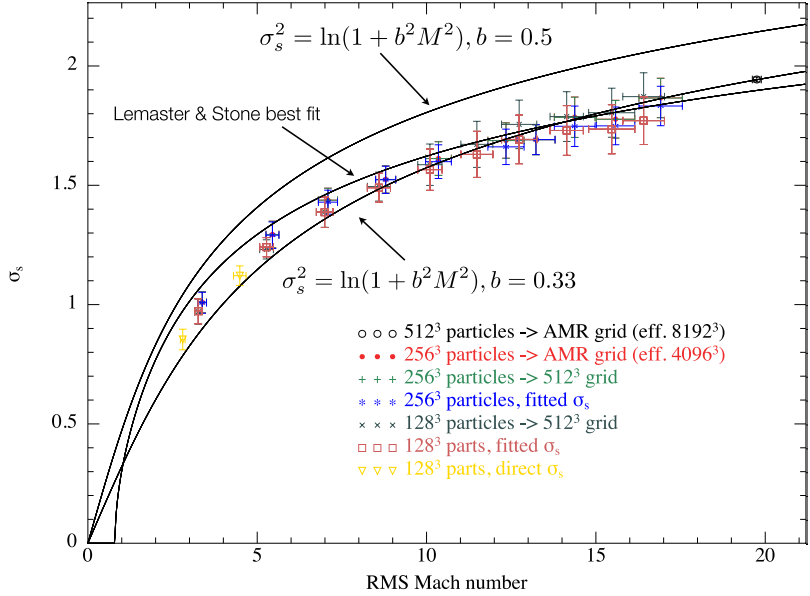} 
 \caption{The measured relationship (points) between the standard deviation in the log density, $\sigma_{s}$ and the RMS Mach number, $\mathcal{M}$, from a series of calculations of solenoidally-driven, supersonic turbulence at various Mach numbers and at different resolutions in both SPH particles and on the grid to which the density field is interpolated. Overplotted is the standard relationship, Eq.~\ref{eq:sigmachlog}, with $b=0.33$ (lower) and $b=0.5$ (above). Also shown is the best fit obtained by \citet{ls08} at lower Mach number, Eq.~\ref{eq:ls08}, which differs only slightly from the standard relation with $b=0.33$. Error bars show the $1\sigma$ deviations from the time-averaged values.}
   \label{fig:rhomach}
\end{center}
\end{figure}

\section{The density variance--Mach number relation in supersonic turbulence}
 One of the questions to arise from the turbulence comparison project was the question of exactly how the width of the density Probability Distribution Function (PDF) in driven, supersonic turbulence depends on the Mach number. Early calculations by \citet{pnj97} found/assumed a relationship between the linear density variance $\sigma_{\rho}^{2}$ and the Mach number $\mathcal{M}$ of the form
\begin{equation}
\sigma_{\rho}^{2} = b^{2} \mathcal{M}^{2},
\label{eq:sigmach}
\end{equation}
such that in the log-normal distribution the standard deviation is given by
\begin{equation}
\sigma_{s}^{2} = \ln (1 + b^{2} \mathcal{M}^{2}),
\label{eq:sigmachlog}
\end{equation}
where $\sigma_{s}\equiv \sigma_{\ln \rho}$ is the standard deviation in the \emph{log} of the density and $b$ is a parameter that was determined empirically to be $b \sim 0.5$ \citep*{pnj97}. Later authors have found considerable variation in the measured value of $b$ based on individual calculations performed at a fixed Mach number. Recently \citet{federrathetal10} have also shown that the value of $b$ also depends strongly on particulars of the fourier-space driving, specifically the balance between energy in solenoidal and compressible modes (with $b\approx 0.33$ for purely solenoidal and $b\approx 1$ for purely compressible driving), which they relate physically to whether or not turbulence is driven by compressions or shear flows.

\begin{figure}
\begin{center}
\includegraphics[width=0.8\columnwidth]{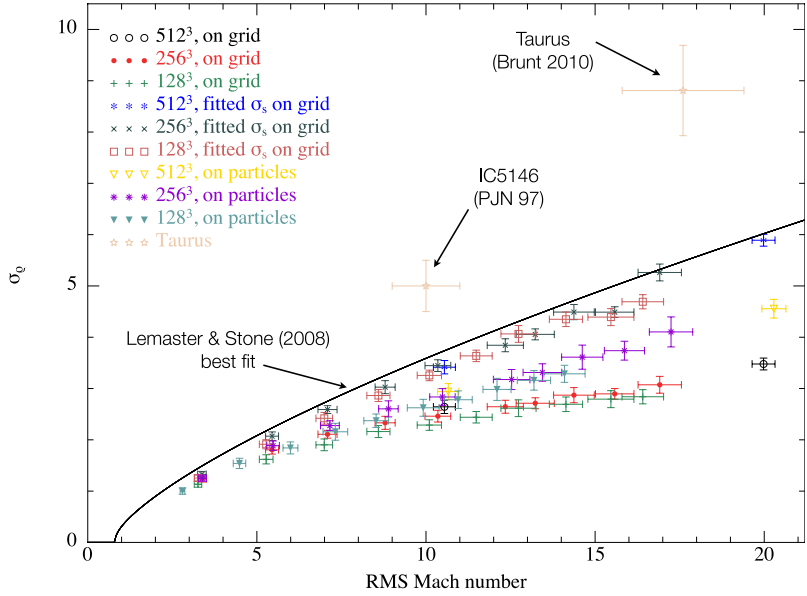} 
 \caption{The directly measured (lower points) and inferred (assuming a log-normal distribution: $\sigma_{\rho}^{2} = \exp(\sigma_{s}^{2}) - 1)$, higher points) relationship between the standard deviation in linear density, $\sigma_{\rho}$, and the RMS Mach number, $\mathcal{M}$, in supersonic turbulence simulations, together with the best-fitting relationship from \citet{ls08} (solid line, similar to $b=0.33$ in Eq.~\ref{eq:sigmach}) and the observational results from IC5146 by \citet{pjn97} and in Taurus from \citet{brunt10}.}
   \label{fig:sigmach-obs}
\end{center}
\end{figure}

 However, relatively few studies have attempted to constrain the relation over a range in Mach number, though the early empirical (but unpublished) work by Padoan et al. was calibrated by such calculations. \citet{ls08} have recently done so, though only in a limited range of Mach numbers (up to $\mathcal{M}\sim 6$), deriving a best-fitting relationship
\begin{equation}
\sigma_{s}^{2} = -0.72 \ln \left(1 + 0.5 \mathcal{M}^{2}\right) + 0.20.
\label{eq:ls08}
\end{equation}
 However we have recently developed a technique for inferring the true 3D variance $\sigma_{\rho}$ from observational data \citep{bfp10a} that, together with the measured Mach number, indicate $b\approx 0.5 \pm 0.05$ in the Taurus Molecular Cloud at Mach 20 \citep{brunt10}, similar to an earlier result derived at in IC5146 at Mach~10 by \citet*{pjn97}. In order to make comparisons with observations it is therefore necessary to pin down the relationship up to these Mach numbers.
 
  We have thus performed a series of calculations with RMS Mach numbers in the range $1-20$, the results of which are shown in Fig.~\ref{fig:rhomach} in terms of $\sigma_{s}$, measured directly as the (volume-weighted) standard deviation in the (gridded) log density field with no assumptions regarding log-normality or otherwise. The difficulty in comparing with observations is that the variance in the \emph{linear} density field is severely truncated by finite resolution effects, but it is this quantity that is measurable observationally (though it will also be underestimated). One way of estimating the true value is to simply assume that the PDF is log-normal, i.e., $\sigma_{\rho}^{2} = \exp (\sigma_{s}^{2}) - 1$. The calculations were observed to approach this relationship as the resolution -- either in terms of the number of particles or the grid to which they are interpolated -- is increased. The standard deviation in the linear density is what is measured in the observational data, a comparison with which is shown in Fig.~\ref{fig:sigmach-obs}. Whilst we find a good fit to Eq.~\ref{eq:sigmachlog} with $b=0.33$ (and also to the LS08 best fit, Eq.~\ref{eq:ls08}) for the purely solenoidal driving we have employed, it is clear that the density variances measured observationally lie significantly above those in the calculations, with $b\approx0.5$ (and this noting that our assumption of log-normality will \emph{overestimate} $\sigma_{\rho}$, if anything). In the heuristic model by \citet{Federrathetal08} this can be explained if the is a significant compressive component to the driving mechanism, or alternatively due to the fact that star-forming molecular clouds are quite obviously self-gravitating which is not accounted for in pure turbulence models.

\section{Conclusions}
In summary:
\begin{itemize}
\item Being a television presenter is easier than getting MHD in SPH to work
\item Development of MHD in SPH would proceed faster in the absence of arguments over basic misunderstandings of SPH which require lengthy comparison projects to resolve.
\item Simulations with supercritical field strengths already tell us that magnetic fields can significantly change star formation even with weak fields.
\item SPH and grid codes agree very well on the statistics of supersonic turbulence provided comparable resolutions are used (for power spectra this means $n_{cells} \approx n_{particles}$, but SPH was found to be much better at resolving dense structures).
\item We have constrained the density variance -- Mach number relation in supersonic turbulence to $\sigma_{s}^{2} = \ln( 1 + b^{2} \mathcal{M}^{2})$, with $b\approx0.33$ up to Mach~20, but observed density variances suggest $b\approx0.5$, higher than can be produced with purely solenoidally-driven turbulence -- rather, some form of compressive driving or additional physics such as gravity is required to explain the observations.
\end{itemize}

\bibliography{sph,mhd,starformation}

\end{document}